\documentclass[
aps,
prl,
10pt,
notitlepage,
nofootinbib,
tightenlines,
floatfix,
twocolumn,
superscriptaddress]{revtex4-2}

\usepackage[english]{babel}
\usepackage{makecell}
\usepackage[margin=1in]{geometry}
\usepackage[colorlinks=true, allcolors=blue]{hyperref}
\hypersetup{
    colorlinks=true,       
    linkcolor=red,          
    citecolor=magenta,        
    filecolor=magenta,      
    urlcolor=cyan,           
    runcolor=cyan
}
\usepackage{bm,bbm,amssymb,amsmath,amsfonts,amsthm,mathrsfs,times}
\usepackage{natbib}
\usepackage{graphicx}
\usepackage{amsmath}
\usepackage{multirow}
\usepackage{color}
\usepackage{bbold}
\usepackage[utf8]{inputenc}
\usepackage{braket}
\usepackage{blkarray}
\usepackage{titlesec}
\usepackage[normalem]{ulem}

\newcommand{\tr}{\mathrm{Tr}}
\newcommand{\pvec}[1]{\vec{#1}\mkern2mu\vphantom{#1}}

\newcommand{\red}[1]{{\color{red} {#1}}}

\graphicspath{ {figures/} }

\begin{document}

\title{Classical Non-Markovian Noise in Symmetry-Preserving Quantum Dynamics}
\author{William M. Watkins}
\email[]{wwatki11@jhu.edu}
\affiliation{William H. Miller III Department of Physics \& Astronomy, 
Johns Hopkins University, Baltimore, Maryland 21218, USA}

\author{Gregory Quiroz}
\email[]{Gregory.Quiroz@jhuapl.edu}
\affiliation{William H. Miller III Department of Physics \& Astronomy,
Johns Hopkins University, Baltimore, Maryland 21218, USA}
\affiliation{Johns Hopkins University Applied Physics Laboratory,
Laurel, Maryland 20723, USA}
\begin{abstract}

In quantum dynamics, symmetries are vital for identifying and assessing conserved quantities that govern the evolution of a quantum system. When promoted to the open quantum system setting, dynamical symmetries can be negatively altered by system-environment interactions, thus, complicating their analysis. Previous work on noisy symmetric quantum dynamics has focused on the Markovian setting, despite the ubiquity of non-Markovian noise in a number of widely used quantum technologies. In this Letter, we develop a framework for quantifying the impact of non-Markovian noise on symmetric quantum evolution via root space decompositions and the filter function formalism. We demonstrate analytically that symmetry-preserving noise maintains the symmetric subspace, while nonsymmetric noise leads to highly specific leakage errors that are block diagonal in the symmetry representation. We support our findings with numerical studies of a transverse-field Ising model and a quantum error detecting code subject to spatiotemporally correlated multiaxis noise. Our results are broadly applicable, providing new analytic insights into the control and characterization of open quantum system dynamics.
\end{abstract}

\maketitle

Symmetries play a crucial role in understanding quantum dynamics, both for simplifying calculations and for gaining deeper insights into their behavior.
In many cases, an exact calculation of the dynamics is infeasible due to complexity or lack of knowledge, but one can still obtain nontrivial insights based on symmetry considerations.
The crux of symmetry-preserving dynamics is Noether's theorem~\cite{Noether_1971}, which associates every differentiable symmetry of a classical action with a conserved quantity. This extends naturally to quantum dynamics, where symmetries of the Hamiltonian under time evolution lead to observables whose moments are conserved~\cite{Sakurai_Napolitano_2020}.
Furthermore, symmetries partition the Hilbert space into invariant subspaces of the time-evolution operator that constrain the dynamics of quantum states~\cite{lastres2024nonuniversalityconservedsuperoperatorsunitary}.
Symmetry considerations play a critical role in 
quantum error correcting codes (QECCs)~\cite{Gottesman_1996, Zanardi_2000, Liu_2023}, quantum key distribution~\cite{Scarani_2009}, adiabatic quantum computing~\cite{Schaller_2010}, many-body physics~\cite{Schmitz2020, vanleeuwen2024revealingsymmetriesquantumcomputing}, quantum information~\cite{Bartlett_2007, Blume_kohout_2010, Marvian_2014}, and variational quantum algorithms~\cite{Hadfield_2019, Shaydulin_2021,tsvelikhovskiy2024sym}. 

In systems subject to environmental influences, generalizations of Noether's theorem provide a framework for understanding the role of symmetry and conservation laws in open quantum systems~\cite{albert_2014, Lostaglio_2017, Albert_2019}.
While dissipative systems rarely admit conserved quantities, symmetries have been key in studying monotones (i.e., functions of a quantum state that decay monotonically)~\cite{Marvian_2014, Styliaris_2020} as well as identifying both decoherence-free subspaces~\cite{lidar_1998, kempe_2001, Fortunato_2002} and uniformly decaying subspaces~\cite{suri2024uniformlydecayingsubspaceserror}. Thus far, studies of symmetries in open quantum systems have focused on Markovian dynamics (i.e., memoryless environments) expressed via the Lindblad master equation and quantum dynamical semigroups~\cite{gorini_kossakowski_sudarshan_1976, lindblad_1976}, where symmetries arise in the Lindblad generator~\cite{Alicki_2007, Baumgartner_2008}. Such investigations extended to the non-Markovian regime remain underexplored, despite their relevance to experimental systems~\cite{Breuer_2016, de_vega_2017, Krantz_2019, burkard2023semi, wintersperger2023neutral, prakash_2024}.

Here, we develop a formalism to quantify the impact of classical, non-Markovian noise in quantum dynamics with symmetries. Our approach draws on concepts from Lie theory and quantum control theory---in particular, root space decompositions and the filter function formalism (FFF)~\cite{Kofman2001universal, cywinski2008fff, Green_2013, Paz_Silva_2017, cerfontaine_2021_prl, cerfontaine_2021_prr}---to identify and characterize the dynamical propagation of noise through the system. We show that noise which preserves the symmetries maintains the symmetric subspace, while symmetry-breaking noise leads to limited, highly specific leakage outside the subspace. Our results apply broadly to any unitary evolution subject to classical non-Markovian noise~\cite{van_Exter_2001, Chenu_2017, Zou2024}. They offer a symmetry-based framework for assessing the impact of noise, identifying characteristic noise decay channels, and devising techniques for combating noise in quantum evolution.

\emph{Faulty quantum evolution}---Consider a quantum system of dimension $N$ with a Hilbert space $\mathcal{H}_S=\mathbb{C}^N$.
Let $\mathcal{B}(\mathcal{H_S})$ be the set of bounded linear operators on $\mathcal{H}_S$ and $\rho\in\mathcal{B}(\mathcal{H_S})$ denote the density operator.
The noiseless system dynamics are generated by a Hamiltonian 
$H_0(t)\in\mathcal{B}(\mathcal{H})$ with the superoperator propagator $\mathcal{U}_0(T,0){[\cdot]}=\mathcal{T}_+e^{-i\int^T_0 dt [H_0(t),\cdot]}$.
This study focuses on quantum systems that, either naturally or by embedding in a higher dimensional Hilbert space, have linearly independent commuting symmetries $\{Q_i\}$, such that
\begin{equation}
    [Q_i,H_0(t)]=0, \forall t.
\end{equation}
These symmetries span an abelian subalgebra $\mathfrak{q}=\rm{span}[\{Q_i\}]$.
Furthermore, let the initial state $\rho(0)$ be an eigenvector of the symmetries $Q\in\mathfrak{q}$, i.e., $Q\rho(0) = q\rho(0)$ for $q\in\lambda(Q)$, where $\lambda(Q)$ is the spectrum of $Q$.
Then it is straightforward to show that the noiseless evolution occurs within a \emph{symmetry-preserving subspace} (SPS), the $q$-eigenspace, i.e., $Q\rho_0(T)=q\rho_0(T)$ for any time $T$, where $\rho_0(T)=\mathcal{U}_0(T,0)[\rho(0)]$ represents the time-evolved state with respect to the noiseless dynamics.

Here, we consider classical non-Markovian noise models that either maintain or break the symmetries. We focus on classical temporally correlated noise models well motivated in superconducting qubits \cite{Bylander_2011, Krantz_2019}, semiconductor qubits \cite{Chatterjee_2021, Connors_2022}, and trapped ions \cite{Soare_2014, Frey_2020}. Namely, the faulty dynamics are described by $H(t) = H_0(t) + H_E(t)$. The error Hamiltonian $H_E(t)$ describes a classical stochastic field coupled to the quantum system~\cite{crow2014noise}:
\begin{equation}
    H_E(t) = \sum_\mu \beta_\mu(t) N_\mu,
\end{equation}
 where $N_\mu$ 
 are traceless bounded linear operators on $\mathcal{H}_S$.
 $\beta_\mu(t)$ are assumed to be wide-sense stationary  Gaussian stochastic processes with zero mean, 
 $\overline{\beta_\mu(t)}=0$, 
 and with a noise power spectrum density (PSD), $S_{\mu\nu}(\omega)=\mathcal{F}_{\omega}\Big[\overline{\beta_\mu(t)\beta_\nu(0)}\Big]$, defined in terms of the two-point correlation function $\overline{\beta_\mu(t)\beta_\nu(0)}$. $\overline{\cdots}$ denotes the statistical ensemble average and $\mathcal{F}_{\omega}$ is the Fourier transform. 
 We call the noise model \emph{symmetry-preserving} with respect to $\mathfrak{q}$ if $[\mathfrak{q},H_E(t)]=0$; thus, $[Q,N_\mu]=0$ $\forall \mu, Q\in\mathfrak{q}$. For \emph{symmetry-breaking} noise, noise operators do not commute with the symmetry group and thus cause transitions out of the SPS. 

\emph{Root space decomposition}---Lie theory is integral in the study of quantum control~\cite{Dirr_2008, Dong_2010} and multiqubit quantum dynamics~\cite{Yen_2021, Fontana_2024, Qvarfort_2025,Ragone_2024, fazio2025cartanquantummetrology, prudhoe2025cartancovariantquantumchannelsppt2}.
We use Lie theory to construct a symmetry-informed operator basis for the error dynamics.
Consider the irreducible representation of $\mathfrak{q}$,
\begin{equation}
    \label{eqn:hilbert_space}
    \mathcal{H}_S = \bigoplus_{\vec q\in\lambda(\{Q_i\})}\mathcal{V}^{(\vec q)},
\end{equation}
which decomposes the Hilbert space into a sum of subspaces invariant under the symmetry group generated by $\mathfrak{q}$. These subspaces are indexed by a tuple of eigenvalues, $\vec q$, corresponding to operators $\{Q_i\}$ which span the subalgebra $\mathfrak{q}$.

Evolution operators acting on 
$\mathcal{H}_S$ form the special unitary group $SU(N)$. This is generated by the special unitary Lie algebra $\mathfrak{g}=\mathfrak{su}(N)$.
An abelian subalgebra, known as the Cartan subalgebra $\mathfrak{h}$, can be constructed for $\mathfrak{g}$ with respect to the symmetries such that the abelian subalgebra $\mathfrak{q}\in\mathfrak{h}$. This is ensured as all $\mathfrak{h}$ are isomorphic. 
The Cartan subalgebra can be used to construct a \emph{root space decomposition} of the Lie algebra $\mathfrak{g} = \mathfrak{h} \bigoplus_{\alpha\in\Phi}\mathfrak{g}_\alpha,$
where the subalgebras $\mathfrak{g}_\alpha$ are indexed by the roots $\alpha$ in the root system $\Phi$. These are defined as
\begin{equation}
    \mathfrak{g}_\alpha = \big\{x \in\mathfrak{g}\mid [h,x]=\alpha(h)x \textrm{ for all } h\in\mathfrak{h}\big\},
\end{equation}
and have the property $\dim\mathfrak{g}_\alpha=1$.
The root system represents the ladder operators of the Cartan subalgebra, which increment the quantum number of $h$ by $\alpha(h)$ for an eigenvector of $\mathfrak{h}$ in $\mathcal{H}_S$. 
The root system $A_{N-1}$ describes $\mathfrak{su}(N)$, meaning that the roots live in the Euclidean space $\alpha\in\mathbb{R}^{N-1}$~\cite{Humphreys_1972}. 
The root space decomposition for $\mathfrak{su}(4)$ is given as an example in the Supplemental Material~\cite{supp}.

One can partition the root system into projection and ladder operators of the symmetry quantum numbers.
We denote the ordered set of eigenvalues as $\vec q\in\lambda(\mathfrak{q})$, i.e., $Q_i x = (\vec q)_i x$ for $x\in\mathfrak{g}^{(\vec q)}$.
Then,
\begin{equation}
    \label{eqn:decomposition}
    \mathfrak{g} = \bigoplus_{\vec q\in\lambda(\mathfrak{q})} \mathfrak{h}^{(\vec q)}\oplus \mathfrak{g}^{(\vec q)}\bigoplus_{\vec q,\pvec q'\in\lambda(\mathfrak{q})}\mathfrak{g}^{(\vec q\to \pvec q')} ,
\end{equation}
where $\mathfrak{h}^{(\vec q)}$ is the projection of the Cartan subalgebra into the $\vec q$ eigenspace, $\mathfrak{g}^{(\vec q)}$ consists of ladder operators that preserve the eigenvalues $\vec q$ and $\mathfrak{g}^{(\vec q\to \vec q')}$ consists of ladder operators that transition the eigenvalues from $\vec q$ to $\pvec{q}'$. 
The first term of Eq.~(\ref{eqn:decomposition}) represents the algebras on each subspace of Eq.~(\ref{eqn:hilbert_space}).
This decomposition is comparable to the total-spin operators $\ket{j,m}\bra{j',m'}$, indexed by the $J^2$ and $J_z$ quantum numbers, generalized to any set of abelian operators.
This eigenspace decomposition is rather powerful and enables general, analytic statements to be made about the dynamics of a stochastic quantum system with symmetries $\{Q_i\}$.
By expanding the dynamics in terms of symmetry-informed operators, we can physically and intuitively understand how errors affect conserved quantities of the system.

\emph{Effective error dynamics}---We assess the impact of noise on the system via the FFF.
The noisy dynamics are isolated from the ideal dynamics using the reverse interaction picture~\cite{youssry2020characterization,quiroz2021quantifying, zhou2022quantum} in the Liouvillian representation, $\mathcal{U}(T,0) = \tilde{\mathcal{U}}_E(T,0)\circ\mathcal{U}_0(T,0)$, where 
$\mathcal{U}(T,0)[\cdot]=
\mathcal{T}_+e^{-i\int^T_0 dt [H(t),\cdot]}$ is the total-time evolution superoperator, and $\tilde{\mathcal{U}}_E(T,0)$ is the superoperator generated by 
$\tilde{H}_E(t)=\mathcal{U}_0(T,t)[H_E(t)]$.
Then the final quantum state averaged over noise becomes
$\overline{\rho(T)} = \overline{\tilde{\mathcal{U}}_E(T)}[\rho_0(T)]$.
We use a cumulant expansion~\cite{kubo_1962} $\overline{\tilde{\mathcal{U}}_E(T)}=e^{\mathcal{C}(T)}$ to derive an effective model for the total error dynamics.
The Liouvillian FFF~\cite{cerfontaine_2021_prl, cerfontaine_2021_prr} is utilized due to its convenient analytic properties, such that $\mathcal{C}(T)[\cdot]=\sum_k \frac{(-i)^k}{k!}\mathcal{C}^{(k)}(T)[\cdot]$ is a superoperator expansion.
We enforce a weak noise assumption, $\|H_E(t)\|T \ll 1$, and truncate the expansion to second order.
The first nonzero terms in the cumulant expansion are
\begin{equation}
    \label{eqn:l_fff_cumulant}
    \mathcal{C}(T) = \sum_{ij} \Big( \chi_{ij}^{(1)}(T) \mathcal{A}_{ij} + \chi_{ij}^{(2)}(T) \mathcal{B}_{ij} \Big),
\end{equation}
where $\mathcal{A}_{ij}=[x_i,[x_j,\cdot]]$ and {$\mathcal{B}_{ij}=[[x_i,x_j],\cdot]$ are superoperators. The coefficients $\chi^{(1)}_{ij}(T)=\sum_{\mu,\nu}\int^\infty_{0} \frac{d\omega}{2\pi}S_{\mu\nu}(\omega)\mathcal{F}^{\mu\nu}_{ij}(\omega,T)$ and $\chi^{(2)}_{ij}(T)=\sum_{\mu,\nu}\int^\infty_{0} \frac{d\omega}{2\pi}S_{\mu\nu}(\omega)\mathcal{G}^{\mu\nu}_{ij}(\omega,T)$ are the decoherence parameters, or spectral overlaps. 
The filter functions (FFs) are $\mathcal{F}^{\mu\nu}_{ij}(\omega,T)=\int^T_0 dt_1 \int^T_0 dt_2 e^{-i\omega(t_1-t_2)} R^\mu_i(t_1)R^\nu_j(t_2)$ and $\mathcal{G}^{\mu\nu}_{ij}(\omega,T)=\int^T_0 dt_1 \int^{t_1}_0 dt_2 e^{-i\omega(t_1-t_2)} R^\mu_i(t_1)R^\nu_j(t_2)$, respectively.
The summation is over the generators of $SU(N)$, i.e., $x_i,x_j\in\mathfrak{g}$, and $R^\mu_i(t)=\tr\big[\mathcal{U}_0(T,t)[x_i] x_\mu\big]$ is the control matrix, 
arising from the frame transform of $H_E(t)$~\cite{Green_2013}. 
{The cumulant operator [Eq.~\eqref{eqn:l_fff_cumulant}] contains two terms, generating dissipative and coherent evolution, respectively~\cite{cerfontaine_2021_prl, cerfontaine_2021_prr}. It can be shown that the former shares similarities with the Lindblad master equation---leading to Markovian dynamics when the noise PSD is flat [i.e, $S_{\mu\nu}(\omega)=S_{\mu\nu}$]---while the latter is uniquely generated by the presence of noise correlations~\cite{supp}.}

\emph{Control matrix decomposition}---The spectral overlaps are determined by $R^\mu_i(t)$ elements which expand $\tilde{H}_E(t)$ in some basis.
Usually the FFs (and control matrices) are calculated with the $n$-qubit Pauli group~\cite{Green_2013, Paz_Silva_2019}. 
This is an ill-informed choice for noisy quantum systems with symmetry.
Instead, we use the generators derived from the eigenspace decomposition 
to define a well-informed basis for expansion and calculation of the FFs.

A key result of our study is that 
$R^\mu_i(t)$ is block diagonal if and only if calculated in the eigenspace decomposition with respect to $\mathfrak{q}$.
We call this the $\mathfrak{q}$-basis.
The control matrix elements are nonzero if and only if $x_i,x_\mu\in\mathfrak{g}$ share the same eigenvalues $\vec q$, or are transitions between the same two eigenvalues of $\vec q$,
that is,
\begin{equation}
    \label{eqn:switching_function}
    R_i^\mu(t) = 0 \mid x_i\in\mathfrak{h}^{(\vec q_1)}\oplus\mathfrak{g}^{(\vec q_1)},x_\mu\in\mathfrak{h}^{(\vec q_2)}\oplus\mathfrak{g}^{(\vec q_2)}.
\end{equation}
for $\vec q_1\neq \vec q_2$.
Since the transformation is generated by $H_0(t)$,
the rotating-frame operator $\mathcal{U}_0(T,t)[x_i]$ has the same root projection $\alpha(\mathfrak{q})$ as $x_i$.
Consequently, the subalgebras $\mathfrak{h}^{(\vec q)}\oplus\mathfrak{g}^{(\vec q)},\mathfrak{g}^{(\vec q\to \pvec q')}$ are invariant under the frame transformation, preserving their orthogonal properties.
Equation~(\ref{eqn:switching_function}) follows from this orthogonality.
The control matrix is exponentially large for a system composed of qubits, so this key result identifies the set of FFs identically zero; thus, leading to a significant reduction in FF calculations needed to model a non-Markovian open quantum system.

\emph{Error channel characterization}---We use the properties of the FFs and root system to examine the impact of symmetry-preserving and symmetry-breaking noise on the
quantum state dynamics. 
First, consider noise that preserves the symmetry of $H_0$, then $S_{\mu\nu}(\omega)\neq 0$ only for $N_\mu,N_\nu\in\mathcal{Z}(\red{\mathfrak{q}})$, the centralizer of the symmetry subalgebra. 
Since the FFs are block diagonal in the $\mathfrak{q}$-basis and the noise is symmetric, the noise perturbation is also block diagonal:
$
    \mathcal{C}(T)[\mathfrak{h}^{(\vec q)}\oplus\mathfrak{g}^{(\vec q)}] \subseteq\mathfrak{h}^{(\vec q)}\oplus\mathfrak{g}^{(\vec q)}.
$
Therefore, by application of $\mathcal{C}(T)$, the error operator preserves the SPS for any noise power or time, i.e., 
\begin{equation}
    \overline{\mathcal{U}_E(T)}[\rho_0(T)] = e^{\mathcal{C}(T)}[\rho_0(T)]\in\mathfrak{h}^{(\vec q)}\oplus\mathfrak{g}^{(\vec q)}.
    \label{eq:sps-evol}
\end{equation}
Albeit intuitive, this result highlights the utility of this theoretical framework to parse the impact of symmetry-preserving noise in open quantum systems.

Next, we consider a noise model including operators that do not commute with $\mathfrak{q}$. 
As such, $N_\mu\in\mathfrak{h}^{(\vec q)}\oplus\mathfrak{g}^{(\vec q)},\mathfrak{g}^{(\vec q\to \pvec q')}$ can be operators within a $\mathfrak{q}$ eigenspace or $\mathfrak{q}$ transitions. We assume no correlations between the symmetric and nonsymmetric noise modes, i.e., $S_{\mu\nu}(\omega)=0$ for $\mathcal{N}_\mu\in\mathfrak{h}^{(\vec q)}\oplus\mathfrak{g}^{(\vec q)}$ and $\mathcal{N}_\nu\in\mathfrak{g}^{(\vec q\to \pvec q')}$.
In this case, we find that
\begin{equation}
    \mathcal{C}(T)[\mathfrak{h}^{(\vec q)}\oplus\mathfrak{g}^{(\vec q)}] \subseteq\bigoplus_{\pvec q'\in\lambda(\{Q_i\})}\Big( \mathfrak{h}^{(\pvec q')}\oplus\mathfrak{g}^{(\pvec q')} \Big),
    \label{eq:non-sps-evol}
\end{equation}
so
\begin{equation}
    \overline{\mathcal{U}_E(T)}[\rho_0(T)]\in\bigoplus_{\pvec q'\in\lambda(\{Q_i\})}\Big( \mathfrak{h}^{(\pvec q')}\oplus\mathfrak{g}^{(\pvec q')} \Big),
\end{equation} 
which implies that the noise perturbation and noise-averaged density operator are block diagonal in the symmetry representation. 
The nonsymmetry preserving noise induces probability amplitude outside the initial symmetry subspace $\mathcal{V}^{(\vec q)}$, but there is no coherence between the eigenspaces.
Importantly, this result provides key insight into the FFs that are most relevant to the dynamical evolution. 
Normally, one needs $\mathcal{O}(N^4)$ FFs, but with this construction, one only needs $\mathcal{O}(N_{\mathfrak{q}}^4)$ elements to characterize classical non-Markovian noise, where $N_{\mathfrak{q}} = \dim Z({\mathfrak{q}})$ and usually $N_{\mathfrak{q}} \ll N$~\cite{tsvelikhovskiy2024sym}.
This powerful, nonintuitive result is demonstrated by the example below. First, we will showcase the utility of this formalism in assessing the dynamics of the system.

\emph{Weak-noise bounds}---Consider the distance between the ideal and noise-averaged 
faulty
density operator, $D = \frac{1}{2}\| (\overline{\mathcal{U}}_E(T,0)-\mathcal{U}_0(T,0))[\rho(0)]\|_1$. In the weak-noise regime ($S_0 T \ll 1$), 
where $S_0=\max S_{\mu\nu}(\omega)$,
we make use of the Dyson expansion
such that
$D \approx \frac{1}{2}\| \mathcal{C}(T)[\rho_0(T)] \|_1$.
From this expression, we can compute an upper bound on the distance via bounds on the cumulants.
Because of the block diagonal nature of Eq.~({\ref{eqn:switching_function}), the bound decouples symmetry-preserving and symmetry-breaking noise, 
\begin{equation}
    \begin{split}
        D\leq \underbrace{\sum_{ij\mu\nu\in\mathfrak{h}^{(\vec q)}\oplus\mathfrak{g}^{(\vec q)}} \psi_{ij}^{\mu\nu}(T) }_{\text{symmetric noise}} + \underbrace{\sum_{\pvec q'\in\lambda(\mathfrak{q})}\sum_{ij\mu\nu\in\mathfrak{g}^{(\vec q\to \pvec q')}} \psi_{ij}^{\mu\nu}(T) }_{\text{nonsymmetric noise}}
    \end{split}
\end{equation}
where this includes an overlap integral between the FFs and noise power spectra, $\psi_{ij}^{\mu\nu}(T) = \int^\infty_0 \frac{d\omega}{2\pi}S_{\mu\nu}(\omega)\big( |\mathcal{F}_{ij}^{\mu\nu}(\omega,T)|
+ |\mathcal{G}_{ij}^{\mu\nu}(\omega,T)|\big)$. 
Note the contrast between $\psi^{\mu\nu}_{ij}(T)$, which sums the magnitude of FFs,
and the signed quantities, $\chi^{(1)}_{ij}(T)$ and $\chi^{(2)}_{ij}(T)$.
{Since the second term sums over the entire spectrum of $\mathfrak{q}$, the bound on the distance increases significantly with the introduction of nonsymmetric noise.}

\begin{figure}[tb]
  \centering
  \includegraphics[width=\linewidth]{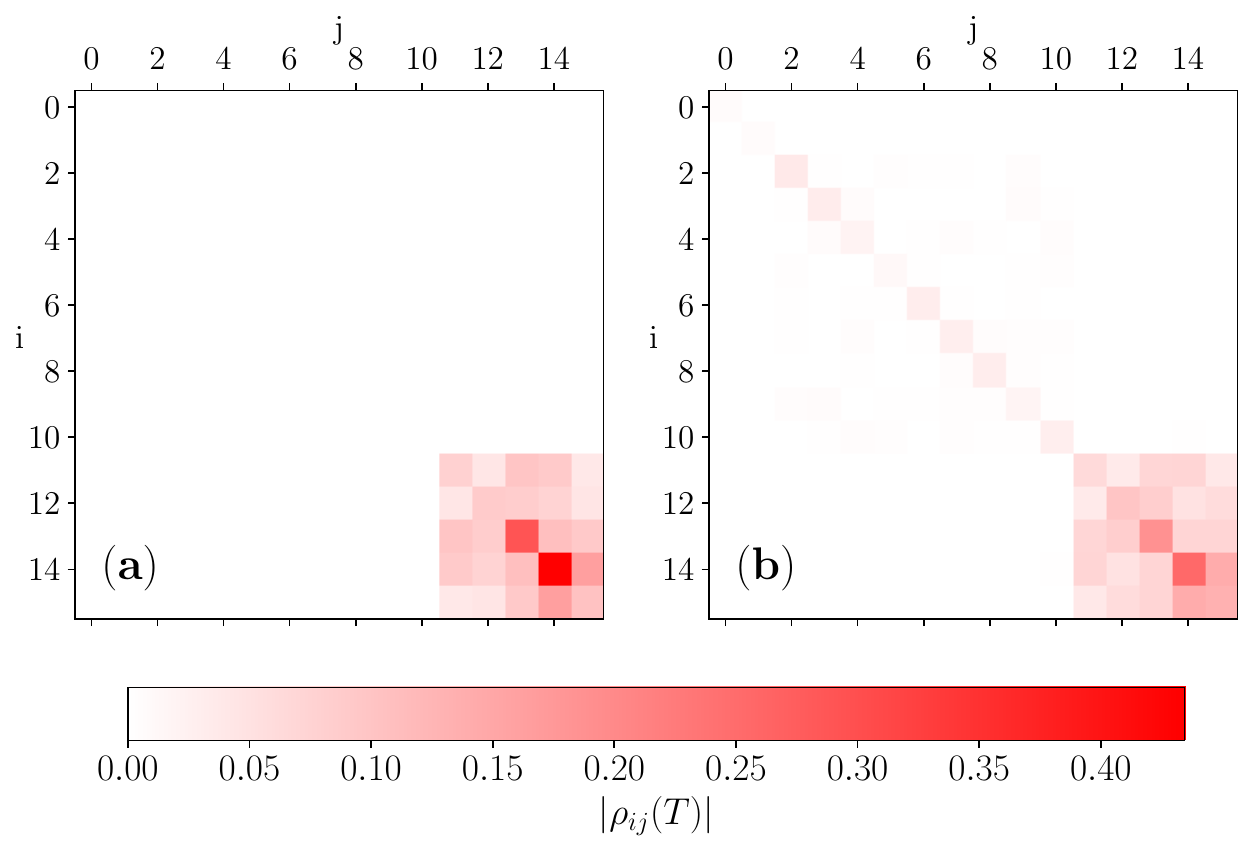}
\caption{ 
    Magnitude of the density matrix elements in the {$\mathfrak{q}$}-basis for the TFIM {with four spins}, averaged over the pink noise ensemble with (a) global dephasing noise and (b) local dephasing noise.
    {The symmetry representation decomposes the Hilbert space into three sectors: $j=\{0, 1, 2\}$.}
    Panel (a) demonstrates that symmetry-preserving noise causes decoherence exclusively in the SPS.
    In contrast, in panel (b), local dephasing leads to specific transitions out of the SPS, constrained by symmetry. The averaged faulty quantum state is block diagonal in the symmetry representation.
    Each calculation is averaged over an ensemble of 20,000 noise trajectories.
    The simulation time is $T \approx 2\tau$, where $\tau$ is the noise correlation length.
    The correlation length was found by fitting an exponential curve to the autocorrelation function $C(t)=\mathcal{F}^{-1}_t[S(\omega)]$.
}
\label{fig:density_matrices}
\end{figure}

\emph{Symmetry-preserving quantum spin dynamics}---We support the above theoretical findings with an example. We consider the evolution generated by the transverse-field Ising model (TFIM),
\begin{equation}
    H_0(t) = \sum_{ij} J_{ij} \sigma^z_i \sigma^z_j + h \sum_{i} \sigma^x_i
    \label{eq:tfim}
\end{equation}
where $h$ and $J_{ij}$ represent local bias and couplings, respectively. $\sigma^\mu_i$ denotes the $\mu$ Pauli operator for the $i$th spin-1/2 particle. 
The TFIM is widely used in material simulations~\cite{georgescu2014sim, heyl2018dynamical} and serves as a foundation for adiabatic quantum computation~\cite{albash_2018} and the quantum approximate optimization algorithm~\cite{farhi2014quantum, farhi2018classification}. 
{While the TFIM has a $\mathbb{Z}_2$ symmetry, if the spin-coupling $J_{ij}$ is an all-to-all uniform interaction, then the
evolution generated by Eq.~(\ref{eq:tfim}) has a stronger symmetry: total angular momentum $J^2$.
While $\mathbb{Z}_2$ has only two eigenvalues ($\pm 1$), 
$J^2$ has $n$ eigenvalues, meaning there are $n$ blocks in the control matrix, making for a more interesting and complex problem.
The symmetry subalgebra is one-dimensional, $\mathfrak{q}=\mathrm{span}[\{J^2 - (\tr[J^2]/N)I\}]$, and conserves total spin $\vec q = (j)$.
The offset is necessary to make it a subalgebra of the traceless $\mathfrak{su}(N)$~\cite{Humphreys_1972}.}
With an initial state $\ket{+}^{\otimes n}$, the ideal dynamics occur with the eigenspace defined by $j=n/2$, also known as the symmetric subspace~\cite{Jiang_2017}.

In Fig.~\ref{fig:density_matrices}, we examine the dynamics of the Ising model {with $J_{ij}=J$} in the presence of dephasing generated by $H_E(t)=\sum^{n}_{i=1}\beta_i(t) \sigma^z_i$ 
{for four spins. While we focus on the $n=4$ results due to computational constraints, the same qualitative behavior is observed for larger spin systems.}
{We distinguish between global dephasing, where the noise is spatially correlated [$\beta_i(t)=\beta(t)$ for all $i$] and local dephasing where the noise is independent at each site.
We use a pink noise model [$S(\omega)\sim 1/\omega$], commonly observed in solid-state and atomic systems~\cite{Bylander_2011, Connors_2022, Yan_2016, Yoneda_2017},
with IR and UV cutoffs ensuring integrability.
While we choose pink noise here, the specifics of the PSD does not affect the algebraic structure of the open quantum system.

In the global dephasing noise model, the noise Hamiltonian simplifies to $H_E(t)=n\beta(t)J_z$, which commutes with the total spin operator and conserves total spin.
} %
Hence, the spin dynamics are restricted to the SPS of the Hilbert space and decohere within that subspace. 
On the other hand, for a local dephasing noise model, we get leakage out of the SPS into the other {eigenspaces}.
This result is generalizable to any $J^2$ preserving quantum dynamics in accordance with Eqs.~(\ref{eq:sps-evol}) and (\ref{eq:non-sps-evol}).

\begin{figure}[tb]
  \centering
  \includegraphics[width=\linewidth]{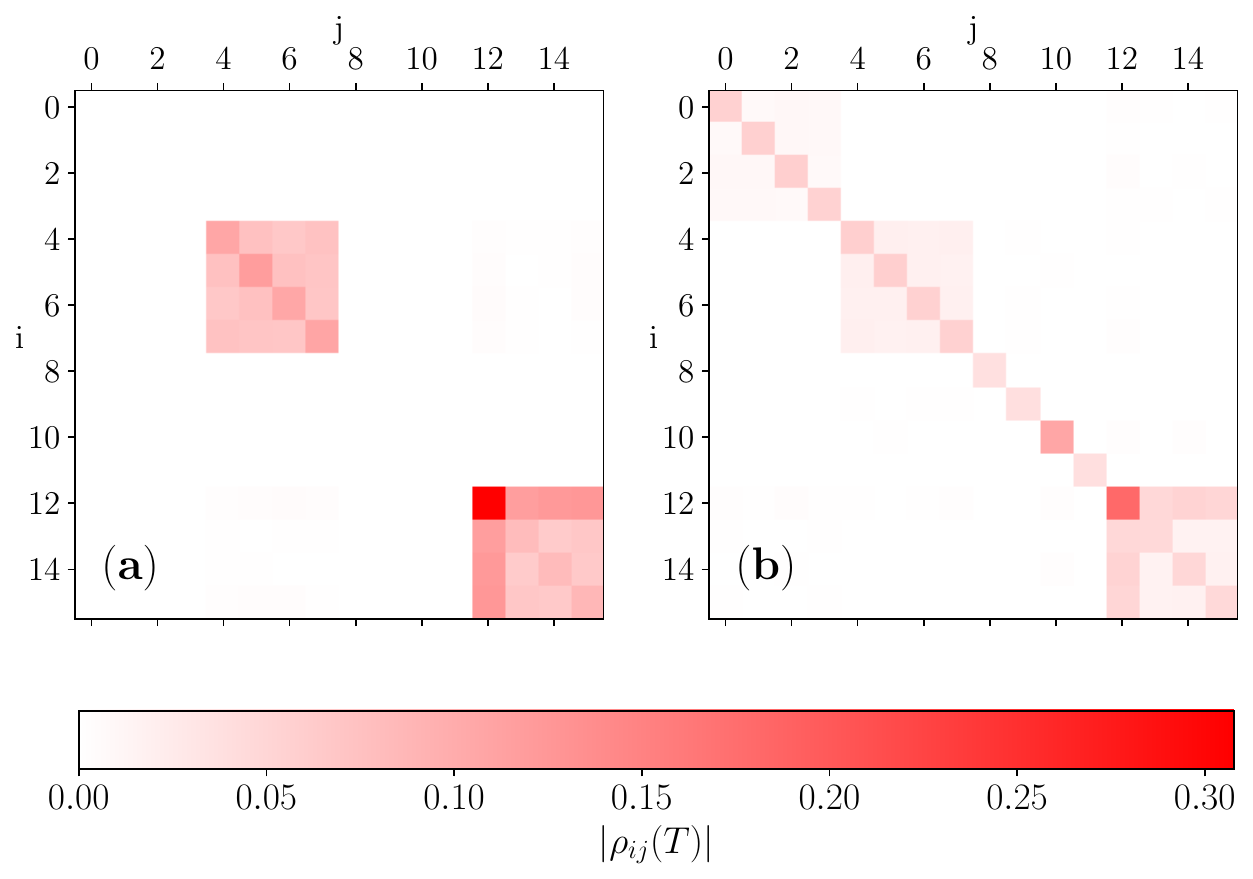}
\caption{ 
    {Magnitude of the density matrix elements in the $\mathfrak{q}$-basis for the $[[4,2,2]]$ code, averaged over the pink noise ensemble with (a) $X$ noise and (b) multiaxis noise.
    Panel (a) demonstrates that the model preserving the $X$ parity causes decoherence within the symmetric subspace and transitions flipping the Z parity.
    In contrast, in panel (b), the model breaks both symmetries, which leads to transitions out of the symmetric subspace into the other three eigenspaces.}
}
\label{fig:error_detecting_code}
\end{figure}

{
\emph{Quantum error-detecting code}---Next, we demonstrate the developed framework with an example relevant to QECCs.
The $[[4,2,2]]$ quantum error-detecting code encodes two logical qubits using four physical qubits~\cite{Vaidman_1996, Leung_1997, Grassl_1997}.
It is the smallest stabilizer code to detect a single-qubit error.
The $[[4,2,2]]$ code does not require active feedback, which has made it amenable to many near-term applications~\cite{Linke_2017, urbanek2020qed, gowrishankar2025logicalerrorrates422encoded, Pokharel2024, vezvaee2025demonstrationhighfidelityentangledlogical}.

The encoded Hamiltonian commutes with the stabilizers, $\{X^{\otimes 4},Z^{\otimes 4}\}$, and the logical states span the eigenspace $\vec{q}_L = (+1,+1)$}.
A single-qubit error is detected when one of these parities is flipped to $-1$.
This problem is more nuanced than the first example, because this symmetry subalgebra is two-dimensional, $\mathfrak{q}=\textrm{span}[\{X^{\otimes 4},Z^{\otimes 4}\}]$.
We construct a Cartan subalgebra  with $\mathfrak{q}$, and use the root space decomposition to formulate a symmetry representation of the operator space.

In Fig.~\ref{fig:error_detecting_code}, we examine the simulated dynamics of the $[[4,2,2]]$ code.
A logical state is encoded and freely evolved and after time $T$, the density operator is measured in the symmetry basis (pre-post-selection).
We inject temporally correlated single- and multiaxis noise into the dynamics, generated by $H_E(t) = \sum^n_{i=1}\beta_i^x(t)\sigma^x_i$ and $H_E(t) = \sum^n_{i=1}\big(\beta_i^x(t)\sigma^x_i + \beta_i^z(t)\sigma^z_i \big)$,
respectively. We choose a pink noise model ($S_{ii}(\omega)\sim 1/\omega$) with no spatial correlations ($S_{ij}(\omega)=0$ for $i\neq j$).
The first noise model breaks the $Z^{\otimes 4}$ symmetry, so there is leakage from the symmetric subspace into the subspace with $-1$ parity.
The second noise model breaks both stabilizers, so there is leakage from the symmetric subspace into the other three eigenspaces.
As predicted by our analytic model, in the $\mathfrak{q}$-basis, the density operator is block diagonal, and the logical FFs and errors are defined by
$e^{\mathcal{C}(T)}[\rho_0(T)]\in \mathfrak{h}^{(\vec{q}_L)}\oplus \mathfrak{g}^{(\vec{q}_L)}$. 
This result has broad impact for future QECC studies, as we may use this framework to inform error characterization, recovery, and logical gate design.
}

\emph{Conclusions}---We leverage root systems and the FFF to analyze the impact of non-Markovian noise on symmetry-preserving quantum dynamics. 
Our approach enables the characterization of noise processes that are either symmetrypreserving or symmetry breaking. 
To this end, we construct novel generators of $\mathfrak{su}(N)$ informed by the symmetries of the ideal system.
We take advantage of the group structure of these generators via the root system to make broad symmetry constraints on open quantum system dynamics.

We demonstrate that decoherence is restricted to the SPS for symmetry-preserving correlated noise, whereas symmetry-breaking correlated noise causes specific leakage outside the SPS. 
{Therefore, we can identify the subalgebra causing temporally correlated logical errors.}
Our results provide key insights into the propagation of noise in symmetric quantum evolution that can be exploited for the design of noise characterization and protection strategies.    

{Our analysis of classical non-Markovian noise in symmetric systems extends beyond the Markovian Lindblad description often used in recent work on symmetries in open quantum systems.}
{While we focus exclusively on classical stochastic processes in this study, the algebraic structure and framework should generalize to stochastic quantum baths. 
We leave that for a future study.}

\begin{acknowledgments}
\emph{Acknowledgments}---The authors would like to thank Yasuo Oda, Nathaniel Watkins, Kevin Schultz, and Leigh Norris for useful discussions and feedback. This work was supported by the U.S. Department of
Energy, Office of Science, Office of Advanced Scientific
Computing Research, Accelerated Research in Quantum
Computing under Awards No. DE-SC0020316 and No. DE-SC0025509.
In addition, this material is based upon work supported by the National Science Foundation under Grant No. 2515049.  Any opinions, findings, and conclusions or recommendations expressed in this material are those of the author(s) and do not necessarily reflect the views of the National Science Foundation.
\end{acknowledgments}

\emph{Data Availability}---—The datasets generated and/or analyzed from this study are not publicly available due to institutional requirements but may be obtained from the authors on reasonable request, pending completion of the necessary review procedures.

\end{document}